\documentclass[%
 aip,
 amsmath,amssymb,
preprint,%
 reprint,%
]{revtex4-1}

\usepackage{xcolor}
\usepackage{graphicx}% Include figure files
\usepackage{dcolumn}% Align table columns on decimal point
\usepackage{bm}% bold math
\usepackage[utf8]{inputenc}
\usepackage[T1]{fontenc}
\usepackage{mathptmx}
\usepackage{etoolbox}

\makeatletter
\def\@email#1#2{%
 \endgroup
 \patchcmd{\titleblock@produce}
  {\frontmatter@RRAPformat}
  {\frontmatter@RRAPformat{\produce@RRAP{*#1\href{mailto:#2}{#2}}}\frontmatter@RRAPformat}
  {}{}
}%
\makeatother
\begin{document}

\preprint{AIP/123-QED}

\title{28 THz soliton frequency comb in a continuous-wave pumped  fiber Fabry-Pérot resonator}
%\title{28 THz cavity soliton in a high quality factor fiber Fabry-Pérot resonnator under continous wave pumping}

% Force line breaks with \\
\author{T. Bunel}
 \email{thomas.bunel@univ-lille.fr}
\author{M. Conforti}%
\author{Z. Ziani}%
\affiliation{%
University of Lille, CNRS, UMR 8523-PhLAM Physique des Lasers, Atomes et Molécules, F-59000, Lille, France%\\This line break forced% with \\
}%

\author{J. Lumeau}
\author{A. Moreau}
\affiliation{ 
Aix Marseille University, CNRS, Centrale Marseille, Institut Fresnel, Marseille, France%\\This line break forced with \textbackslash\textbackslash
}%

\author{A. Fernandez}
\author{O. Llopis}
\author{G. Bourcier}
 \altaffiliation[Also at ]{CNES, 18 Avenue Edouard Belin, F-31401 Toulouse, France}%Lines break automatically or can be forced with \\
\affiliation{%
LAAS-CNRS, Université de Toulouse, CNRS, 7 avenue de Colonel Roche, 31031 Toulouse, France%\\This line break forced% with \\
}%

\author{A. Mussot$^1$}
 %\altaffiliation
 
\date{\today}% It is always \today, today,
             %  but any date may be explicitly specified

\begin{abstract}
We report the generation of an optical frequency comb featuring 28 THz bandwidth, sustained by a single 80 fs  cavity soliton recirculating in a fiber Fabry-Pérot resonator. %, resulting in a frequency comb
This large spectrum is comparable to frequency combs obtained with microresonators operating in the anomalous dispersion regime. Thanks to the compact design and the easy coupling of the resonator, cavity solitons can be generated in an all-fiber experimental setup with a continuous wave pumping scheme. We also observe the generation  of a dispersive wave at higher frequencies which is supported by  higher-order dispersion. These observations align remarkably well with both numerical simulations and the established theory of cavity solitons.
\end{abstract}

\maketitle

%\begin{quotation}
%The ``lead paragraph'' is encapsulated with the \LaTeX\ 
%\verb+quotation+ environment and is formatted as a single paragraph before the first section heading. 
%(The \verb+quotation+ environment reverts to its usual meaning after the first sectioning command.) 
%Note that numbered references are allowed in the lead paragraph.
%
%The lead paragraph will only be found in an article being prepared for the journal \textit{Chaos}.
%\end{quotation}

\section{Introduction}

Nonlinear Kerr cavities have enabled the generation of cavity solitons (CSs) \cite{leo_temporal_2010, herr_temporal_2014,obrzud_temporal_2017,kippenberg_dissipative_2018,pasquazi_micro-combs_2018}, which offer precise femtosecond sources and wide-ranging optical frequency combs (OFCs) with repetition rates spanning from MHz to THz, impacting a wide range of cross-disciplinary applications: data transmission \cite{marin-palomo_microresonator-based_2017} and processing \cite{tan_rf_2021}, ranging \cite{trocha_ultrafast_2018}, microwave photonics \cite{lucas_ultralow-noise_2020}, dual-comb spectroscopy \cite{suh_microresonator_2016}, and astronomical spectrograph calibration \cite{obrzud_microphotonic_2019}. These solitons arise as localised solutions of the Lugiato–Lefever equation\cite{Lugiato1987,Haelterman1992} and can be observed in resonators with high-quality factors. The emergence of CSs relies on the double balance between anomalous group velocity dispersion (GVD) and  Kerr nonlinearity on one side, and between losses and energy injection (typically achieved through continuous-wave (CW) laser pumping) on the other side. %  of the waveguide and the periodic boundary conditions imposed by the cavity, typically achieved through continuous-wave (CW) laser pumping. 
Owing to their high-quality factor and compact design (cavity length of hundreds of microns), microresonators have attracted significant attention over the last decade \cite{pasquazi_micro-combs_2018,kippenberg_dissipative_2018,sun_applications_2023}. Despite these impressive performances, launching and collecting light in these resonators can be challenging, requiring advanced fiber coupling devices such as a prism fiber taper \cite{cai_observation_2000} or advanced coupling methods for chip microresonators \cite{razzari_cmos-compatible_2010}, and while progresses on packaging are on going, it is still an obstacle for fiber applications. Another way to generate OFCs in resonators consists in using all-fiber ring cavities of tens of meters in length \cite{leo_temporal_2010,jang_observation_2014}, whose effective quality factor can reach several millions by including an amplifier within the cavity \cite{englebert_temporal_2021}. Spectra obtained using these resonators architecture extend over several THz, almost like microresonators, but they have two major drawbacks. Firstly, the line spacing is in the MHz range, which limits the field of application (mostly in the GHz range \cite{sun_applications_2023}), and secondly, they are not compact. An interesting alternative consists in taking benefit of fiber Fabry–Pérot (FFP) resonators of several centimeters in length. They are a good compromise between fiber ring cavities and microresonators, offering several tens of millions of Q-factors, as well as easy connection to photonic devices with a standard physical-contact fiber connector (FC/PC) and small size \cite{obrzud_temporal_2017,jia_photonic_2020,xiao_near-zero-dispersion_2023,bunel2023observation,li_ultrashort_2022,postdeadline}. CS generation has already been demonstrated with these devices using either a pulsed pumping scheme \cite{obrzud_temporal_2017} or stabilization management through the Brillouin effect \cite{jia_photonic_2020,nie_synthesized_2022}. These recent studies have paved the way for this novel method of OFC generation. However, the generated CSs via CW pumping still have durations exceeding 200~fs \cite{jia_photonic_2020,nie_synthesized_2022}, which falls short of the performance achieved by microresonators. 

This study demonstrates that manufacturing a FFP resonator using a highly nonlinear fiber with low GVD at the pump wavelength enables the generation of sub-100 fs CSs. Moreover, we implemented an advanced triggering experimental setup enabling an accurate and easy control of the detuning to explore the different regimes of the cavity before reaching the soliton states together with an efficient stabilization feature. Additionally, the inherent low GVD characteristics, combined with the large spanning of the generated CS, leads to the emergence of dispersive radiation due to higher-order dispersion, which permits to observe a broad spectrum spanning over 28 THz.

\section{Fabrication of the cavity}

\begin{figure}
\includegraphics{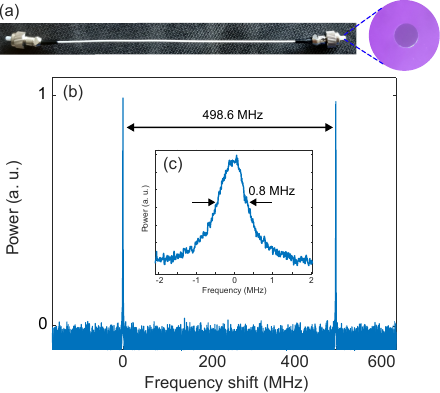}% Here is how to import EPS art
\caption{\label{fig:cavite} Description of the FFP resonator. (a) Photograph of device with one of its deposited mirror. (b) Transmission function of resonator in linear regime. (c) Zoom on a cavity resonance}
\end{figure}

The FFP cavity used for the study is depicted in Fig.~\ref{fig:cavite}. It is made from an optical highly nonlinear fiber (HNLF Thorlabs-HN1550P) of length L~=~20.63~cm, a group velocity dispersion (GVD) of $\beta_2 = -0.8$~ps$^2$km$^{-1}$, a third-order dispersion (TOD) of $\beta_3 = 0.03$~ps$^3$km$^{-1}$ at the pump wavelength (1550~nm) and a nonlinear coefficient of $\gamma = 10.8$~W$^{-1}$km$^{-1}$. Both fiber ends are mounted in FCs/PC, and Bragg mirrors are deposited at each extremity with a physical vapor deposition technique, to achieve $99.86\%$ reflectance over a 100 nm bandwidth \cite{zideluns_automated_2021}. Fig.~\ref{fig:cavite}(a) shows a connector with its deposited mirror. The linear transfer function is shown in Fig.~\ref{fig:cavite}(b) and (c). This architecture leads to a resonator with a $0.8$~MHz linewidth resonance at full width half maximum, a linear coupling efficiency of $25~\%$ and a peak resonance transmission of $5~\%$. The free spectral range (FSR) is measured at $498.6$~MHz, resulting in a finesse $\mathcal{F}=620$, and a quality-factor $Q=230\cdot10^6$. One of the great advantages of this FFP cavity with respect to a microresonator \cite{cai_observation_2000,razzari_cmos-compatible_2010} is its plug-and-play feature into an all-fiber photonic device.

\section{Experimental setup}

\begin{figure}
\includegraphics{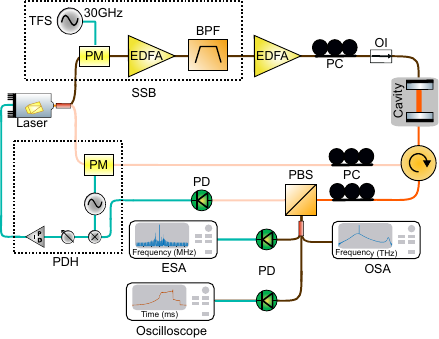}
\caption{\label{fig:montage} Experimental setup with a two-arms stabilization system. Brown line: nonlinear beam; beige line: control beam. Both beams are perpendicularly polarized to each other. TFS: Tunable Frequency Synthesizer; PM: Phase Modulator; EDFA: Erbium Doped Fiber Amplifier; PC: Polarization Controller; OI: Optical Isolator; PD: Photodiode; PBS: Polarization Beam Splitter; PDH: Pound-Drever-Hall; SSB: single-side-band generator.}
\end{figure}

The FFP resonator is exploited in the experimental setup described in Fig~\ref{fig:montage}. The generation of CSs requires bistable operation and specific excitation protocols \cite{pasquazi_micro-combs_2018,sun_applications_2023}. One of the most popular and efficient solution consists in performing a scan of the resonance from blue to red and to stop at a precise laser frequency, which fixes a specific cavity detuning. To achieve precise control over the cavity detuning, a two-arms stabilization scheme is implemented \cite{nishimoto_thermal_2022,li_ultrashort_2022}. The CW laser is split into two beams: one beam serves as the control beam for stabilizing the laser on a cavity resonance (control beam), while the other beam acts as the pump beam for the cavity (nonlinear beam). To allow independent handling of the beams and prevent their interaction within the cavity, we take advantage of the natural birefringent of optical fibers. The polarization states of the control and nonlinear beams are crossed polarized, along the two main polarization axes of the fiber cavity by means of polarization controllers. They are separated at the output using a polarization beam splitter [see PBS in Fig.~\ref{fig:montage}]. The stabilization process [lower beige arm in Fig.~\ref{fig:montage}] is achieved through a Pound-Drever-Hall (PDH) system, enabling laser locking at the top of cavity resonance, and with the main interest to be insensitive to amplitude variations \cite{drever_laser_1983,black_introduction_2001}. On the other hand, the detuning of the nonlinear beam [upper brown arm in Fig.~\ref{fig:montage}], is controlled using a homemade tunable single-sideband generator [see SSB in Fig.~\ref{fig:montage}]. The nearest side band of a modulated beam, obtained with a phase modulator driven by a $30$~GHz tunable frequency synthesizer (TFS), is isolated to obtain a pump signal with a tunable frequency shift. This approach allows the nonlinear beam frequency to experience similar variations to those of the control beam. It also makes possible to adjust the frequency shift between the two, and consequently to control the detuning value of the nonlinear beam, by simply modifying the value of the TFS frequency. Thanks to the TFS frequency ramp function, it is possible to scan cavity resonances, or manually change the TFS frequency and therefore the detuning value. The nonlinear beam is further amplified by an erbium-doped fiber amplifier to reach a power of 1~W before being launched into the cavity. However, the SMF-HNLF transitions, at the cavity input and output, induce important losses due to the difference in effective area overlap which is estimated to $3$~dB. Thus, the effective pump power at the cavity input is estimated at $0.5$~W.

\section{Characterization of the different nonlinear regimes}

\begin{figure*}
\includegraphics{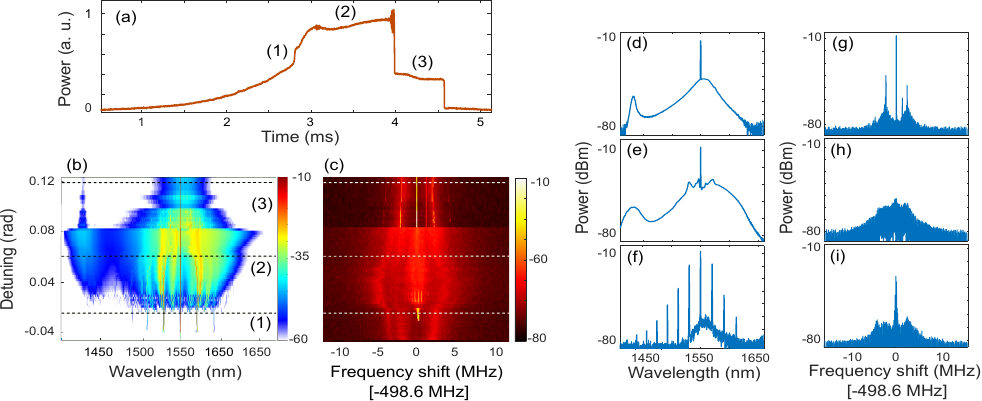}
\caption{\label{fig:dynamique}Experimental recordings as a function of the cavity detuning. (a): Nonlinear transfer function with a fast redshift scan of the cavity (scan speed = 2~GHz/s). (b): Spectrum evolution with cavity detuning. (c): RF beatnote evolution with cavity detuning. (d), (e) and (f): Snapshots of each comb structure spectrum, corresponding to the black dashed lines in (b). (g), (h) and (i): Snapshots of each comb structure RF beatnote, corresponding to the white dashed lines in (c). (f) and (i): MI, $\delta=-0.015$~rad. (e) and (h): chaos, $\delta=0.06$~rad. (d) and (g): soliton, $\delta=0.12$~rad. The video in the supplemental material provides a comprehensive overview of the comb structure variation throughout the scan (Multimedia view).}
\end{figure*}

Thanks to this setup, we can easily observe the distinct nonlinear regimes of the cavity, varying with the detuning, through the use of a basic CW pump. It is worth noting that the use of a PDH system for stabilization makes the experimental setup very robust compared to the use of a simple Proportional-Integral-Derivative (PID) systems. It allows to compensate environmental vibrations and the thermal variation. First, a classic fast redshift scan is applied. The TFS frequency is sweeped from 30~GHz to 30,5~GHz with a speed of 2~GHz/s. The evolution of the output power through the scan is recorded thanks to a photodiode and an oscilloscope, and is represented in Fig.~\ref{fig:dynamique}(a). As expected, we observe three different regions [(1), (2), (3) in Fig.~\ref{fig:dynamique}(a)], corresponding to the different well-known comb structures in Kerr resonators, in sequence: modulation instability (MI), chaos and CSs \cite{pasquazi_micro-combs_2018,kippenberg_dissipative_2018,herr_temporal_2014,coen_universal_2013,parra-rivas_dynamics_2014}. Second, in a way to observe the evolution of these three nonlinear regimes, 
%we perform a slow redshift scan recording 
we manually change the detuning value (i.e. TFS frequency value), recording the optical spectrum and radio frequency (RF) beatnote centered at the first beatnote (498.6~MHz), of the generated signal, with an optical spectrum analyzer (OSA) and an electrical spectrum analyzer (ESA), respectively. Fig.~\ref{fig:dynamique}(b) and (c) illustrate the evolution of the experimental generated signals as a function of the detuning, which can be obtained by the relation: $\Delta\delta=-\frac{4\pi n L}{c}\Delta\nu$, where $\Delta\delta$ and $\Delta\nu$ are the detuning variation and TFS frequency variation, respectively (c and n are speed of light in vacuum and the effective refractive index of the fiber mode). In these figures, the three nonlinear regimes can easily be identified with a clear separation between each. Fig.~\ref{fig:dynamique}(d)-(i) show several characteristic examples of the spectrum and RF beatnote of the three comb structures to get a clearer insight. (1)~MI comb formation, characterized by its symmetric sidelobes around the pump in the spectral domain [Fig.~\ref{fig:dynamique}(b) and (f)], resulting to a stable oscillation as the RF spectrum shows [Fig.~\ref{fig:dynamique}(c) and (i)]. (2)~MI lobe mixing leads to a chaotic transmission variation and produces a chaotic comb [Fig.~\ref{fig:dynamique}(b) and (e)]. The chaotic regime is well illustrated by a huge broadening of the beatnote as showed in Fig.~\ref{fig:dynamique}(c) and (h). The spectrum broadens, and a spectral component appears at 1430~nm due to the TOD as we will discuss below [Fig.~\ref{fig:dynamique}(b) and (e)]. (3)~CSs are generated resulting in a broad coherent OFC over 200~nm (i.e. 28~THz) [Fig.~\ref{fig:dynamique}(d) and (g)]. Interestingly, Fig.~\ref{fig:dynamique}(b) shows the existence of different CSs regime, indicating the circulation of multiple solitons within the cavity. However, the sensitivity of the detection system in Fig.~\ref{fig:dynamique}(a) does not allow to clearly highlight the different soliton regimes, from several CSs to a single one. The low sensitivity of the used photodiode might be the reason behind this discrepancy, as multiple steps may be present but not detectable. 

\begin{figure}
\includegraphics{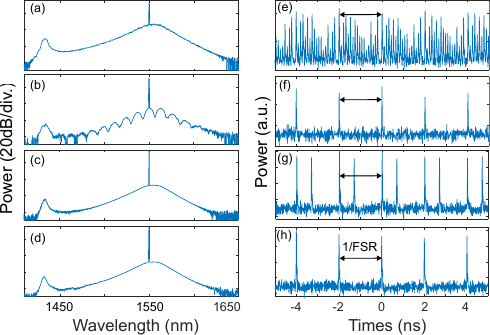}% Here is how to import EPS art
\caption{\label{fig:FVST}Comparison of different signals representing in the spectral and time domain. (a)-(d): Spectral domain measurements with an OSA. (e)-(h): Corresponding time domain measurements with a 30~GHz PD and 70~GHz oscilloscope}
\end{figure}

Furthermore, establishing the presence of a single soliton within the resonator is not straightforward, and a seemingly smooth spectrum at the cavity output is not sufficient for confirmation. This is exemplified in Fig.~\ref{fig:FVST}, where both time domain and spectral domain measurements, conducted using an OSA and a $30$~GHz PD combined to a $70$~GHz oscilloscope, respectively, are depicted. In Fig.~\ref{fig:FVST}(a) and (e), the presence of multiple solitons circulating in the cavity results in a scrambled spectrum and a continuous sequence of oscillations in the time domain due to the limitations of the PD in resolving all circulating solitons. When a cluster of solitons propagates, the time trace reveals a single pulse every roundtrip time ($=1/FSR=2.0056$~ns) [Fig.~\ref{fig:FVST}(f)], potentially suggesting a single soliton generation process. However, the periodic modulations of $1.5$~THz in the spectrum [Fig.~\ref{fig:FVST}(b)], indicates several solitons spaced by $600$~fs (=$1/1.5$~THz) are generated at each roundtrip \cite{obrzud_temporal_2017}. They cannot be resolved through the time domain measurements due to the limited bandpass of the detection system ($30$~GHz). Conversely, Fig.~\ref{fig:FVST}(c) and (g) illustrate a different scenario where multiple solitons propagate far apart, resulting in modulations within the spectrum that are too narrow to be resolved in the spectral domain with a common OSA. However, these instances are discernible in the time domain measurement, where several pulses appear each cavity roundtrip time. The only case demonstrating the generation of a single soliton is depicted in Fig.~\ref{fig:FVST}(d) and (h), characterized by a smooth recorded spectrum and a time domain measurement exhibiting a single pulse every roundtrip time at the same time. This comprehensive analysis underscores the system's capability to generate a single soliton within the FFP resonator employing CW pumping.

\begin{figure}
\includegraphics{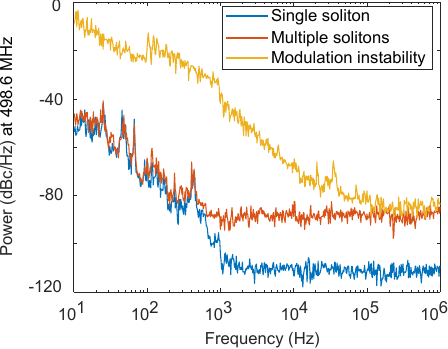}% Here is how to import EPS art
\caption{\label{fig:phaseNoise}Phase noise spectrum measurements of different generated signals. Modulation instability: $\delta=-0.015$~rad; multiple solitons: $\delta=0.1$~rad; single soliton: $\delta=0.12$~rad}
\end{figure}

In order to get a complete characterisation of the dynamics of the system, we record the phase noise spectra corresponding to each regime in Fig.~\ref{fig:phaseNoise}: MI comb, multiple soliton comb, and single soliton comb. These measurements confirm that CSs present the most stable regimes. As expected, the phase noise of the MI comb is significantly higher (40~dB) compared to CSs. An interesting observation is that the phase noise of the multiple soliton comb closely resembles that of the single soliton comb in the low frequencies. However, in the high frequencies, the multiple soliton comb demonstrates considerably higher phase noise compared to the single soliton comb, 30~dB higher, with a comparative level to the MI comb.

\section{Numerical simulations}

\begin{figure*}
\includegraphics{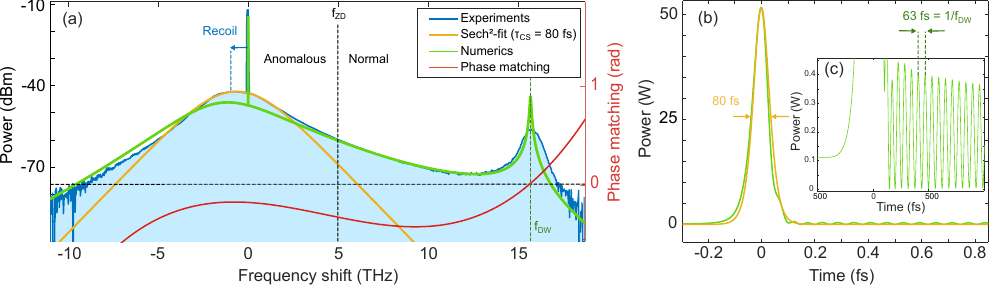}% Here is how to import EPS art
\caption{\label{fig:signals}Cavity soliton in higly nonlinear FFP resonator. (a) Spectral domain.  (b) Time domain. Blue line: experiment; green line: simulation; yellow line: sech² curve fit; red line: phase matching curve. f$_{ZD}$ indicates the frequency of the zero-dispersion wavelength and f$_{DW}$ indicates the frequency of the DW.}
\end{figure*}

We reproduced the experimental results by numerically solving the Lugiato-Lefever equation adapted for FFP cavities (FP-LLE)
\cite{cole_theory_2018,obrzud_temporal_2017,Campbell2023}:
    \begin{align}
  \nonumber  & t_R \frac{\partial \psi}{\partial \tau} = -(\alpha +\delta)\psi  + \theta \sqrt{P_{in}} \\
  &+ 2L \left[ \frac{\beta_3}{6}\frac{\partial^3}{\partial t^3}-i\frac{\beta_2}{2}\frac{\partial^2}{\partial t^2} +i\gamma |\psi|^2 +i\gamma\frac{2}{t_R} \int_{-t_R/2}^{t_R/2}|\psi|^2dt\right] \psi, \label{eq:FP_LLE}
\end{align}
where $\psi$ is the field envelope in units of $\sqrt{W}$, $t$ is the retarded time in the pulse reference frame, $\tau$ is the a slow time counting the number of roundtrips, $P_{in}=0.5$~W is the input power, $t_R=1/FSR$ ($=2.0056$~ns) is the roundtrip time, $\theta=0.0529$ is the transmissivity of the mirror and $\alpha=\pi/\mathcal{F}$ accounts for the total cavity losses (valid for $\mathcal{F}\gg1$). It has recently been demonstrated that this model is well-suited within this parameter range, and its short computational time makes it highly advantageous compared to employing two-coupled nonlinear Schr\"odinger equations \cite{firth_stability_1981, firth_analytic_2021,ziani_theory_2023}. We used the same parameters as in experiments and set the detuning to $\delta=0.12$.  Fig.~\ref{fig:signals}(a) shows the measured soliton spectrum, spanning over an impressive 28~THz. 
The experimental spectrum is in good agreement with the numerical predictions obtained by FP-LLE, represented in green in Fig.~\ref{fig:signals}(a). As for microresonators \cite{yi_soliton_2015}, the use of a fiber with low GVD and a high nonlinear coefficient enables the generation of broader spectra, compared to those achieved with conventional telecom fibers \cite{nie_synthesized_2022,obrzud_temporal_2017}. 
%Note that the SMF-HNLF transition induce 3~dB losses, i.e. an effective input power $P_{in}=0.5$~W. 
The simulation reveals the generation of a soliton of 80~fs duration [green line in Fig.~\ref{fig:signals}(b)], which agrees with the sech²-envelope fit [yellow lines in Fig.~\ref{fig:signals}(a) and (b)]. 
 
In addition, an important spectral peak 15 THz away from the pump is observed. 
This corresponds to  a dispersive wave (DW), also known as Cherenkov radiation, which is emitted by the CS due to the TOD \cite{conforti_dispersive_2013,erkintalo_cascaded_2012,milian_soliton_2014,jang_observation_2014,brasch_photonic_2016,wildi_dissipative_2023}. Indeed, the cavity is pumped only 5 THz away from the zero-dispersion wavelength, allowing an efficient radiation process. The frequency of the DW with respect to the soliton satisfies the phase matching condition \cite{milian_soliton_2014,jang_observation_2014,conforti_dispersive_2013} reported in Eq. (\ref{eq:phaseMatching}): %whose right-hand side corresponds to the red line in Fig.~\ref{fig:signals}(a).
\begin{eqnarray}
\label{eq:phaseMatching}
\frac{2L\beta_3}{6}\omega^3+\frac{2L\beta_2}{2}\omega^2-D\omega-\delta=0.
\end{eqnarray}

Here, $\omega$ is the normalized angular frequency shift of the driving field, and D represents the group-delay accumulated by the temporal CS with respect to the driving field over one round trip (in units of time). Indeed, TOD makes the CS group-velocity slightly different from that of the driving field, leading to a spectral recoil \cite{milian_soliton_2014,jang_observation_2014}. This phenomenon is observable in both the experimental and simulated spectra presented in Fig.~\ref{fig:signals}(a). Thus, the soliton propagates together with the extended radiation tail attached to it [green line in Fig.~\ref{fig:signals}(b) and (c)]. The measured DW frequency position (15.77~THz) is almost identical to the one derived from the phase matching condition (15.69~THz) [red line in Fig.~\ref{fig:signals}(a)], itself identical to the frequency position of the simulated DW [green line in Fig.~\ref{fig:signals}(a)].
The drift delay is calculated as $D=\beta_2\Omega_{CS}2L+\beta_3\Omega_{CS}^2L$, where the recoil frequency shift $\Omega_{CS}/(2\pi) = 1105$~GHz is determined through the numerical simulation \cite{jang_observation_2014} [green line in Fig.~\ref{fig:signals}(a)]. 
It is worth noting that the nonlinear contribution, which depends on the cavity length and CS background power, is not incorporated in the phase matching condition, due to its negligible impact. Nevertheless, it becomes relevant in specific studies \cite{jang_observation_2014}. In the context of FFP cavities, accounting for this would entail considering the specific characteristics of the two-way light circulation, and the additional phase arising from cross-phase modulation \cite{firth_analytic_2021,cole_theory_2018,ziani_theory_2023}. However, this aspect falls outside the scope of this study but presents a potential avenue for exploration in subsequent investigations involving adapted cavities.

\section{Conclusion}

In this study, we have reported the generation of an OFC spanning over 28~THz (i.e. 200~nm), corresponding to a 80~fs cavity soliton duration emitting a dispersive wave, by using a FFP resonator pumped by a CW laser. The highly reflective mirrors and the use of highly nonlinear fiber contribute to achieving a high-quality factor cavity, enhancing nonlinear performance and proving advantageous for Kerr comb generation. The use of this kind of cavity benefits of the ease of implementation into photonic systems by means of its FC/PC connectors. Moreover, our advanced setup with an independent control of the cavity detuning % by means of an homemade single side band generator, 
enabled a smooth tuning of the cavity detuning to observe the dynamics of the system, as well as an excellent stabilization of the cavity with a PDH system to reach -115~dBc/Hz in the best case. The overlap of the CS with the normal dispersion leads to the generation of a DW at 15~THz from the pump. These experimental results are in excellent agreement with numerical simulations. This work contribute to the development of new platforms to generate OFCs, with a view of new applications for fiber systems.

\begin{acknowledgments}
The authors are grateful to Nicolas Englebert and François Leo for fruitful discussions. 

The present research was supported by the agence Nationale de la Recherche (Programme Investissements d’Avenir, I-SITE VERIFICO, FARCO); Ministry of Higher Education and Research; European Regional Development Fund (Photonics for Society P4S),  the CNRS (IRP LAFONI); Hauts de France Council (GPEG project); A.N.R. ASTRID ROLLMOPS; and the university of Lille (LAI HOLISTIC)
\end{acknowledgments}

\section*{Data Availability Statement}

The data that support the findings of this study are available from the corresponding author upon reasonable request.

\nocite{*}
\section*{References}
\bibliography{references}% Produces the bibliography via BibTeX.

%merlin.mbs aipnum4-1.bst 2010-07-25 4.21a (PWD, AO, DPC) hacked
%Control: key (0)
%Control: author (8) initials jnrlst
%Control: editor formatted (1) identically to author
%Control: production of article title (0) allowed
%Control: page (1) range
%Control: year (1) truncated
%Control: production of eprint (0) enabled
\begin{thebibliography}{41}%
\makeatletter
\providecommand \@ifxundefined [1]{%
 \@ifx{#1\undefined}
}%
\providecommand \@ifnum [1]{%
 \ifnum #1\expandafter \@firstoftwo
 \else \expandafter \@secondoftwo
 \fi
}%
\providecommand \@ifx [1]{%
 \ifx #1\expandafter \@firstoftwo
 \else \expandafter \@secondoftwo
 \fi
}%
\providecommand \natexlab [1]{#1}%
\providecommand \enquote  [1]{``#1''}%
\providecommand \bibnamefont  [1]{#1}%
\providecommand \bibfnamefont [1]{#1}%
\providecommand \citenamefont [1]{#1}%
\providecommand \href@noop [0]{\@secondoftwo}%
\providecommand \href [0]{\begingroup \@sanitize@url \@href}%
\providecommand \@href[1]{\@@startlink{#1}\@@href}%
\providecommand \@@href[1]{\endgroup#1\@@endlink}%
\providecommand \@sanitize@url [0]{\catcode `\\12\catcode `\$12\catcode
  `\&12\catcode `\#12\catcode `\^12\catcode `\_12\catcode `\%12\relax}%
\providecommand \@@startlink[1]{}%
\providecommand \@@endlink[0]{}%
\providecommand \url  [0]{\begingroup\@sanitize@url \@url }%
\providecommand \@url [1]{\endgroup\@href {#1}{\urlprefix }}%
\providecommand \urlprefix  [0]{URL }%
\providecommand \Eprint [0]{\href }%
\providecommand \doibase [0]{http://dx.doi.org/}%
\providecommand \selectlanguage [0]{\@gobble}%
\providecommand \bibinfo  [0]{\@secondoftwo}%
\providecommand \bibfield  [0]{\@secondoftwo}%
\providecommand \translation [1]{[#1]}%
\providecommand \BibitemOpen [0]{}%
\providecommand \bibitemStop [0]{}%
\providecommand \bibitemNoStop [0]{.\EOS\space}%
\providecommand \EOS [0]{\spacefactor3000\relax}%
\providecommand \BibitemShut  [1]{\csname bibitem#1\endcsname}%
\let\auto@bib@innerbib\@empty
%</preamble>
\bibitem [{\citenamefont {Leo}\ \emph {et~al.}(2010)\citenamefont {Leo},
  \citenamefont {Coen}, \citenamefont {Kockaert}, \citenamefont {Gorza},
  \citenamefont {Emplit},\ and\ \citenamefont
  {Haelterman}}]{leo_temporal_2010}%
  \BibitemOpen
  \bibfield  {author} {\bibinfo {author} {\bibfnamefont {F.}~\bibnamefont
  {Leo}}, \bibinfo {author} {\bibfnamefont {S.}~\bibnamefont {Coen}}, \bibinfo
  {author} {\bibfnamefont {P.}~\bibnamefont {Kockaert}}, \bibinfo {author}
  {\bibfnamefont {S.-P.}\ \bibnamefont {Gorza}}, \bibinfo {author}
  {\bibfnamefont {P.}~\bibnamefont {Emplit}}, \ and\ \bibinfo {author}
  {\bibfnamefont {M.}~\bibnamefont {Haelterman}},\ }\bibfield  {title}
  {\enquote {\bibinfo {title} {Temporal cavity solitons in one-dimensional kerr
  media as bits in an all-optical buffer},}\ }\href@noop {} {\bibfield
  {journal} {\bibinfo  {journal} {Nature Photonics}\ }\textbf {\bibinfo
  {volume} {4}},\ \bibinfo {pages} {471--476} (\bibinfo {year}
  {2010})}\BibitemShut {NoStop}%
\bibitem [{\citenamefont {Herr}\ \emph {et~al.}(2014)\citenamefont {Herr},
  \citenamefont {Brasch}, \citenamefont {Jost}, \citenamefont {Wang},
  \citenamefont {Kondratiev}, \citenamefont {Gorodetsky},\ and\ \citenamefont
  {Kippenberg}}]{herr_temporal_2014}%
  \BibitemOpen
  \bibfield  {author} {\bibinfo {author} {\bibfnamefont {T.}~\bibnamefont
  {Herr}}, \bibinfo {author} {\bibfnamefont {V.}~\bibnamefont {Brasch}},
  \bibinfo {author} {\bibfnamefont {J.~D.}\ \bibnamefont {Jost}}, \bibinfo
  {author} {\bibfnamefont {C.~Y.}\ \bibnamefont {Wang}}, \bibinfo {author}
  {\bibfnamefont {N.~M.}\ \bibnamefont {Kondratiev}}, \bibinfo {author}
  {\bibfnamefont {M.~L.}\ \bibnamefont {Gorodetsky}}, \ and\ \bibinfo {author}
  {\bibfnamefont {T.~J.}\ \bibnamefont {Kippenberg}},\ }\bibfield  {title}
  {\enquote {\bibinfo {title} {Temporal solitons in optical microresonators},}\
  }\href@noop {} {\bibfield  {journal} {\bibinfo  {journal} {Nature Photonics}\
  }\textbf {\bibinfo {volume} {8}},\ \bibinfo {pages} {145--152} (\bibinfo
  {year} {2014})}\BibitemShut {NoStop}%
\bibitem [{\citenamefont {Obrzud}, \citenamefont {Lecomte},\ and\ \citenamefont
  {Herr}(2017)}]{obrzud_temporal_2017}%
  \BibitemOpen
  \bibfield  {author} {\bibinfo {author} {\bibfnamefont {E.}~\bibnamefont
  {Obrzud}}, \bibinfo {author} {\bibfnamefont {S.}~\bibnamefont {Lecomte}}, \
  and\ \bibinfo {author} {\bibfnamefont {T.}~\bibnamefont {Herr}},\ }\bibfield
  {title} {\enquote {\bibinfo {title} {Temporal solitons in microresonators
  driven by optical pulses},}\ }\href@noop {} {\bibfield  {journal} {\bibinfo
  {journal} {Nature Photonics}\ }\textbf {\bibinfo {volume} {11}},\ \bibinfo
  {pages} {600--607} (\bibinfo {year} {2017})}\BibitemShut {NoStop}%
\bibitem [{\citenamefont {Kippenberg}\ \emph {et~al.}(2018)\citenamefont
  {Kippenberg}, \citenamefont {Gaeta}, \citenamefont {Lipson},\ and\
  \citenamefont {Gorodetsky}}]{kippenberg_dissipative_2018}%
  \BibitemOpen
  \bibfield  {author} {\bibinfo {author} {\bibfnamefont {T.~J.}\ \bibnamefont
  {Kippenberg}}, \bibinfo {author} {\bibfnamefont {A.~L.}\ \bibnamefont
  {Gaeta}}, \bibinfo {author} {\bibfnamefont {M.}~\bibnamefont {Lipson}}, \
  and\ \bibinfo {author} {\bibfnamefont {M.~L.}\ \bibnamefont {Gorodetsky}},\
  }\bibfield  {title} {\enquote {\bibinfo {title} {Dissipative {Kerr} solitons
  in optical microresonators},}\ }\href@noop {} {\bibfield  {journal} {\bibinfo
   {journal} {Science}\ }\textbf {\bibinfo {volume} {361}},\ \bibinfo {pages}
  {eaan8083} (\bibinfo {year} {2018})}\BibitemShut {NoStop}%
\bibitem [{\citenamefont {Pasquazi}\ \emph {et~al.}(2018)\citenamefont
  {Pasquazi}, \citenamefont {Peccianti}, \citenamefont {Razzari}, \citenamefont
  {Moss}, \citenamefont {Coen}, \citenamefont {Erkintalo}, \citenamefont
  {Chembo}, \citenamefont {Hansson}, \citenamefont {Wabnitz}, \citenamefont
  {Del’Haye}, \citenamefont {Xue}, \citenamefont {Weiner},\ and\
  \citenamefont {Morandotti}}]{pasquazi_micro-combs_2018}%
  \BibitemOpen
  \bibfield  {author} {\bibinfo {author} {\bibfnamefont {A.}~\bibnamefont
  {Pasquazi}}, \bibinfo {author} {\bibfnamefont {M.}~\bibnamefont {Peccianti}},
  \bibinfo {author} {\bibfnamefont {L.}~\bibnamefont {Razzari}}, \bibinfo
  {author} {\bibfnamefont {D.~J.}\ \bibnamefont {Moss}}, \bibinfo {author}
  {\bibfnamefont {S.}~\bibnamefont {Coen}}, \bibinfo {author} {\bibfnamefont
  {M.}~\bibnamefont {Erkintalo}}, \bibinfo {author} {\bibfnamefont {Y.~K.}\
  \bibnamefont {Chembo}}, \bibinfo {author} {\bibfnamefont {T.}~\bibnamefont
  {Hansson}}, \bibinfo {author} {\bibfnamefont {S.}~\bibnamefont {Wabnitz}},
  \bibinfo {author} {\bibfnamefont {P.}~\bibnamefont {Del’Haye}}, \bibinfo
  {author} {\bibfnamefont {X.}~\bibnamefont {Xue}}, \bibinfo {author}
  {\bibfnamefont {A.~M.}\ \bibnamefont {Weiner}}, \ and\ \bibinfo {author}
  {\bibfnamefont {R.}~\bibnamefont {Morandotti}},\ }\bibfield  {title}
  {\enquote {\bibinfo {title} {Micro-combs: {A} novel generation of optical
  sources},}\ }\href@noop {} {\bibfield  {journal} {\bibinfo  {journal}
  {Physics Reports}\ }\textbf {\bibinfo {volume} {729}},\ \bibinfo {pages}
  {1--81} (\bibinfo {year} {2018})}\BibitemShut {NoStop}%
\bibitem [{\citenamefont {Marin-Palomo}\ \emph {et~al.}(2017)\citenamefont
  {Marin-Palomo}, \citenamefont {Kemal}, \citenamefont {Karpov}, \citenamefont
  {Kordts}, \citenamefont {Pfeifle}, \citenamefont {Pfeiffer}, \citenamefont
  {Trocha}, \citenamefont {Wolf}, \citenamefont {Brasch}, \citenamefont
  {Anderson}, \citenamefont {Rosenberger}, \citenamefont {Vijayan},
  \citenamefont {Freude}, \citenamefont {Kippenberg},\ and\ \citenamefont
  {Koos}}]{marin-palomo_microresonator-based_2017}%
  \BibitemOpen
  \bibfield  {author} {\bibinfo {author} {\bibfnamefont {P.}~\bibnamefont
  {Marin-Palomo}}, \bibinfo {author} {\bibfnamefont {J.~N.}\ \bibnamefont
  {Kemal}}, \bibinfo {author} {\bibfnamefont {M.}~\bibnamefont {Karpov}},
  \bibinfo {author} {\bibfnamefont {A.}~\bibnamefont {Kordts}}, \bibinfo
  {author} {\bibfnamefont {J.}~\bibnamefont {Pfeifle}}, \bibinfo {author}
  {\bibfnamefont {M.~H.~P.}\ \bibnamefont {Pfeiffer}}, \bibinfo {author}
  {\bibfnamefont {P.}~\bibnamefont {Trocha}}, \bibinfo {author} {\bibfnamefont
  {S.}~\bibnamefont {Wolf}}, \bibinfo {author} {\bibfnamefont {V.}~\bibnamefont
  {Brasch}}, \bibinfo {author} {\bibfnamefont {M.~H.}\ \bibnamefont
  {Anderson}}, \bibinfo {author} {\bibfnamefont {R.}~\bibnamefont
  {Rosenberger}}, \bibinfo {author} {\bibfnamefont {K.}~\bibnamefont
  {Vijayan}}, \bibinfo {author} {\bibfnamefont {W.}~\bibnamefont {Freude}},
  \bibinfo {author} {\bibfnamefont {T.~J.}\ \bibnamefont {Kippenberg}}, \ and\
  \bibinfo {author} {\bibfnamefont {C.}~\bibnamefont {Koos}},\ }\bibfield
  {title} {\enquote {\bibinfo {title} {Microresonator-based solitons for
  massively parallel coherent optical communications},}\ }\href@noop {}
  {\bibfield  {journal} {\bibinfo  {journal} {Nature}\ }\textbf {\bibinfo
  {volume} {546}},\ \bibinfo {pages} {274--279} (\bibinfo {year}
  {2017})}\BibitemShut {NoStop}%
\bibitem [{\citenamefont {Tan}\ \emph {et~al.}(2021)\citenamefont {Tan},
  \citenamefont {Xu}, \citenamefont {Wu}, \citenamefont {Morandotti},
  \citenamefont {Mitchell},\ and\ \citenamefont {Moss}}]{tan_rf_2021}%
  \BibitemOpen
  \bibfield  {author} {\bibinfo {author} {\bibfnamefont {M.}~\bibnamefont
  {Tan}}, \bibinfo {author} {\bibfnamefont {X.}~\bibnamefont {Xu}}, \bibinfo
  {author} {\bibfnamefont {J.}~\bibnamefont {Wu}}, \bibinfo {author}
  {\bibfnamefont {R.}~\bibnamefont {Morandotti}}, \bibinfo {author}
  {\bibfnamefont {A.}~\bibnamefont {Mitchell}}, \ and\ \bibinfo {author}
  {\bibfnamefont {D.~J.}\ \bibnamefont {Moss}},\ }\bibfield  {title} {\enquote
  {\bibinfo {title} {{RF} and microwave photonic temporal signal processing
  with {Kerr} micro-combs},}\ }\href@noop {} {\bibfield  {journal} {\bibinfo
  {journal} {Advances in Physics: X}\ }\textbf {\bibinfo {volume} {6}},\
  \bibinfo {pages} {1838946} (\bibinfo {year} {2021})}\BibitemShut {NoStop}%
\bibitem [{\citenamefont {Trocha}\ \emph {et~al.}(2018)\citenamefont {Trocha},
  \citenamefont {Karpov}, \citenamefont {Ganin}, \citenamefont {Pfeiffer},
  \citenamefont {Kordts}, \citenamefont {Wolf}, \citenamefont {Krockenberger},
  \citenamefont {Marin-Palomo}, \citenamefont {Weimann}, \citenamefont
  {Randel}, \citenamefont {Freude}, \citenamefont {Kippenberg},\ and\
  \citenamefont {Koos}}]{trocha_ultrafast_2018}%
  \BibitemOpen
  \bibfield  {author} {\bibinfo {author} {\bibfnamefont {P.}~\bibnamefont
  {Trocha}}, \bibinfo {author} {\bibfnamefont {M.}~\bibnamefont {Karpov}},
  \bibinfo {author} {\bibfnamefont {D.}~\bibnamefont {Ganin}}, \bibinfo
  {author} {\bibfnamefont {M.~H.~P.}\ \bibnamefont {Pfeiffer}}, \bibinfo
  {author} {\bibfnamefont {A.}~\bibnamefont {Kordts}}, \bibinfo {author}
  {\bibfnamefont {S.}~\bibnamefont {Wolf}}, \bibinfo {author} {\bibfnamefont
  {J.}~\bibnamefont {Krockenberger}}, \bibinfo {author} {\bibfnamefont
  {P.}~\bibnamefont {Marin-Palomo}}, \bibinfo {author} {\bibfnamefont
  {C.}~\bibnamefont {Weimann}}, \bibinfo {author} {\bibfnamefont
  {S.}~\bibnamefont {Randel}}, \bibinfo {author} {\bibfnamefont
  {W.}~\bibnamefont {Freude}}, \bibinfo {author} {\bibfnamefont {T.~J.}\
  \bibnamefont {Kippenberg}}, \ and\ \bibinfo {author} {\bibfnamefont
  {C.}~\bibnamefont {Koos}},\ }\bibfield  {title} {\enquote {\bibinfo {title}
  {Ultrafast optical ranging using microresonator soliton frequency combs},}\
  }\href@noop {} {\bibfield  {journal} {\bibinfo  {journal} {Science}\ }\textbf
  {\bibinfo {volume} {359}},\ \bibinfo {pages} {887--891} (\bibinfo {year}
  {2018})}\BibitemShut {NoStop}%
\bibitem [{\citenamefont {Lucas}\ \emph {et~al.}(2020)\citenamefont {Lucas},
  \citenamefont {Brochard}, \citenamefont {Bouchand}, \citenamefont {Schilt},
  \citenamefont {Südmeyer},\ and\ \citenamefont
  {Kippenberg}}]{lucas_ultralow-noise_2020}%
  \BibitemOpen
  \bibfield  {author} {\bibinfo {author} {\bibfnamefont {E.}~\bibnamefont
  {Lucas}}, \bibinfo {author} {\bibfnamefont {P.}~\bibnamefont {Brochard}},
  \bibinfo {author} {\bibfnamefont {R.}~\bibnamefont {Bouchand}}, \bibinfo
  {author} {\bibfnamefont {S.}~\bibnamefont {Schilt}}, \bibinfo {author}
  {\bibfnamefont {T.}~\bibnamefont {Südmeyer}}, \ and\ \bibinfo {author}
  {\bibfnamefont {T.~J.}\ \bibnamefont {Kippenberg}},\ }\bibfield  {title}
  {\enquote {\bibinfo {title} {Ultralow-noise photonic microwave synthesis
  using a soliton microcomb-based transfer oscillator},}\ }\href@noop {}
  {\bibfield  {journal} {\bibinfo  {journal} {Nature Communications}\ }\textbf
  {\bibinfo {volume} {11}},\ \bibinfo {pages} {374} (\bibinfo {year}
  {2020})}\BibitemShut {NoStop}%
\bibitem [{\citenamefont {Suh}\ \emph {et~al.}(2016)\citenamefont {Suh},
  \citenamefont {Yang}, \citenamefont {Yang}, \citenamefont {Yi},\ and\
  \citenamefont {Vahala}}]{suh_microresonator_2016}%
  \BibitemOpen
  \bibfield  {author} {\bibinfo {author} {\bibfnamefont {M.-G.}\ \bibnamefont
  {Suh}}, \bibinfo {author} {\bibfnamefont {Q.-F.}\ \bibnamefont {Yang}},
  \bibinfo {author} {\bibfnamefont {K.~Y.}\ \bibnamefont {Yang}}, \bibinfo
  {author} {\bibfnamefont {X.}~\bibnamefont {Yi}}, \ and\ \bibinfo {author}
  {\bibfnamefont {K.~J.}\ \bibnamefont {Vahala}},\ }\bibfield  {title}
  {\enquote {\bibinfo {title} {Microresonator soliton dual-comb
  spectroscopy},}\ }\href@noop {} {\bibfield  {journal} {\bibinfo  {journal}
  {Science}\ }\textbf {\bibinfo {volume} {354}},\ \bibinfo {pages} {600--603}
  (\bibinfo {year} {2016})}\BibitemShut {NoStop}%
\bibitem [{\citenamefont {Obrzud}\ \emph {et~al.}(2019)\citenamefont {Obrzud},
  \citenamefont {Rainer}, \citenamefont {Harutyunyan}, \citenamefont
  {Anderson}, \citenamefont {Liu}, \citenamefont {Geiselmann}, \citenamefont
  {Chazelas}, \citenamefont {Kundermann}, \citenamefont {Lecomte},
  \citenamefont {Cecconi}, \citenamefont {Ghedina}, \citenamefont {Molinari},
  \citenamefont {Pepe}, \citenamefont {Wildi}, \citenamefont {Bouchy},
  \citenamefont {Kippenberg},\ and\ \citenamefont
  {Herr}}]{obrzud_microphotonic_2019}%
  \BibitemOpen
  \bibfield  {author} {\bibinfo {author} {\bibfnamefont {E.}~\bibnamefont
  {Obrzud}}, \bibinfo {author} {\bibfnamefont {M.}~\bibnamefont {Rainer}},
  \bibinfo {author} {\bibfnamefont {A.}~\bibnamefont {Harutyunyan}}, \bibinfo
  {author} {\bibfnamefont {M.~H.}\ \bibnamefont {Anderson}}, \bibinfo {author}
  {\bibfnamefont {J.}~\bibnamefont {Liu}}, \bibinfo {author} {\bibfnamefont
  {M.}~\bibnamefont {Geiselmann}}, \bibinfo {author} {\bibfnamefont
  {B.}~\bibnamefont {Chazelas}}, \bibinfo {author} {\bibfnamefont
  {S.}~\bibnamefont {Kundermann}}, \bibinfo {author} {\bibfnamefont
  {S.}~\bibnamefont {Lecomte}}, \bibinfo {author} {\bibfnamefont
  {M.}~\bibnamefont {Cecconi}}, \bibinfo {author} {\bibfnamefont
  {A.}~\bibnamefont {Ghedina}}, \bibinfo {author} {\bibfnamefont
  {E.}~\bibnamefont {Molinari}}, \bibinfo {author} {\bibfnamefont
  {F.}~\bibnamefont {Pepe}}, \bibinfo {author} {\bibfnamefont {F.}~\bibnamefont
  {Wildi}}, \bibinfo {author} {\bibfnamefont {F.}~\bibnamefont {Bouchy}},
  \bibinfo {author} {\bibfnamefont {T.~J.}\ \bibnamefont {Kippenberg}}, \ and\
  \bibinfo {author} {\bibfnamefont {T.}~\bibnamefont {Herr}},\ }\bibfield
  {title} {\enquote {\bibinfo {title} {A microphotonic astrocomb},}\
  }\href@noop {} {\bibfield  {journal} {\bibinfo  {journal} {Nature Photonics}\
  }\textbf {\bibinfo {volume} {13}},\ \bibinfo {pages} {31--35} (\bibinfo
  {year} {2019})}\BibitemShut {NoStop}%
\bibitem [{\citenamefont {Lugiato}\ and\ \citenamefont
  {Lefever}(1987)}]{Lugiato1987}%
  \BibitemOpen
  \bibfield  {author} {\bibinfo {author} {\bibfnamefont {L.~A.}\ \bibnamefont
  {Lugiato}}\ and\ \bibinfo {author} {\bibfnamefont {R.}~\bibnamefont
  {Lefever}},\ }\bibfield  {title} {\enquote {\bibinfo {title} {{Spatial
  dissipative structures in passive optical systems}},}\ }\href {\doibase
  10.1103/PhysRevLett.58.2209} {\bibfield  {journal} {\bibinfo  {journal}
  {Physical Review Letters}\ }\textbf {\bibinfo {volume} {58}},\ \bibinfo
  {pages} {2209--2211} (\bibinfo {year} {1987})}\BibitemShut {NoStop}%
\bibitem [{\citenamefont {Haelterman}, \citenamefont {Trillo},\ and\
  \citenamefont {Wabnitz}(1992)}]{Haelterman1992}%
  \BibitemOpen
  \bibfield  {author} {\bibinfo {author} {\bibfnamefont {M.}~\bibnamefont
  {Haelterman}}, \bibinfo {author} {\bibfnamefont {S.}~\bibnamefont {Trillo}},
  \ and\ \bibinfo {author} {\bibfnamefont {S.}~\bibnamefont {Wabnitz}},\
  }\bibfield  {title} {\enquote {\bibinfo {title}
  {{Additive-modulation-instability ring laser in the normal dispersion regime
  of a fiber.}}}\ }\href {\doibase 10.1364/OL.17.000745} {\bibfield  {journal}
  {\bibinfo  {journal} {Optics letters}\ }\textbf {\bibinfo {volume} {17}},\
  \bibinfo {pages} {745--747} (\bibinfo {year} {1992})}\BibitemShut {NoStop}%
\bibitem [{\citenamefont {Sun}\ \emph {et~al.}(2023)\citenamefont {Sun},
  \citenamefont {Wu}, \citenamefont {Tan}, \citenamefont {Xu}, \citenamefont
  {Li}, \citenamefont {Morandotti}, \citenamefont {Mitchell},\ and\
  \citenamefont {Moss}}]{sun_applications_2023}%
  \BibitemOpen
  \bibfield  {author} {\bibinfo {author} {\bibfnamefont {Y.}~\bibnamefont
  {Sun}}, \bibinfo {author} {\bibfnamefont {J.}~\bibnamefont {Wu}}, \bibinfo
  {author} {\bibfnamefont {M.}~\bibnamefont {Tan}}, \bibinfo {author}
  {\bibfnamefont {X.}~\bibnamefont {Xu}}, \bibinfo {author} {\bibfnamefont
  {Y.}~\bibnamefont {Li}}, \bibinfo {author} {\bibfnamefont {R.}~\bibnamefont
  {Morandotti}}, \bibinfo {author} {\bibfnamefont {A.}~\bibnamefont
  {Mitchell}}, \ and\ \bibinfo {author} {\bibfnamefont {D.~J.}\ \bibnamefont
  {Moss}},\ }\bibfield  {title} {\enquote {\bibinfo {title} {Applications of
  optical microcombs},}\ }\href@noop {} {\bibfield  {journal} {\bibinfo
  {journal} {Advances in Optics and Photonics}\ }\textbf {\bibinfo {volume}
  {15}},\ \bibinfo {pages} {86} (\bibinfo {year} {2023})}\BibitemShut {NoStop}%
\bibitem [{\citenamefont {Cai}, \citenamefont {Painter},\ and\ \citenamefont
  {Vahala}(2000)}]{cai_observation_2000}%
  \BibitemOpen
  \bibfield  {author} {\bibinfo {author} {\bibfnamefont {M.}~\bibnamefont
  {Cai}}, \bibinfo {author} {\bibfnamefont {O.}~\bibnamefont {Painter}}, \ and\
  \bibinfo {author} {\bibfnamefont {K.~J.}\ \bibnamefont {Vahala}},\ }\bibfield
   {title} {\enquote {\bibinfo {title} {Observation of {Critical} {Coupling} in
  a {Fiber} {Taper} to a {Silica}-{Microsphere} {Whispering}-{Gallery} {Mode}
  {System}},}\ }\href@noop {} {\bibfield  {journal} {\bibinfo  {journal}
  {Physical Review Letters}\ }\textbf {\bibinfo {volume} {85}},\ \bibinfo
  {pages} {74--77} (\bibinfo {year} {2000})}\BibitemShut {NoStop}%
\bibitem [{\citenamefont {Razzari}\ \emph {et~al.}(2010)\citenamefont
  {Razzari}, \citenamefont {Duchesne}, \citenamefont {Ferrera}, \citenamefont
  {Morandotti}, \citenamefont {Chu}, \citenamefont {Little},\ and\
  \citenamefont {Moss}}]{razzari_cmos-compatible_2010}%
  \BibitemOpen
  \bibfield  {author} {\bibinfo {author} {\bibfnamefont {L.}~\bibnamefont
  {Razzari}}, \bibinfo {author} {\bibfnamefont {D.}~\bibnamefont {Duchesne}},
  \bibinfo {author} {\bibfnamefont {M.}~\bibnamefont {Ferrera}}, \bibinfo
  {author} {\bibfnamefont {R.}~\bibnamefont {Morandotti}}, \bibinfo {author}
  {\bibfnamefont {S.}~\bibnamefont {Chu}}, \bibinfo {author} {\bibfnamefont
  {B.~E.}\ \bibnamefont {Little}}, \ and\ \bibinfo {author} {\bibfnamefont
  {D.~J.}\ \bibnamefont {Moss}},\ }\bibfield  {title} {\enquote {\bibinfo
  {title} {{CMOS}-compatible integrated optical hyper-parametric oscillator},}\
  }\href@noop {} {\bibfield  {journal} {\bibinfo  {journal} {Nature Photonics}\
  }\textbf {\bibinfo {volume} {4}},\ \bibinfo {pages} {41--45} (\bibinfo {year}
  {2010})}\BibitemShut {NoStop}%
\bibitem [{\citenamefont {Jang}\ \emph {et~al.}(2014)\citenamefont {Jang},
  \citenamefont {Erkintalo}, \citenamefont {Murdoch},\ and\ \citenamefont
  {Coen}}]{jang_observation_2014}%
  \BibitemOpen
  \bibfield  {author} {\bibinfo {author} {\bibfnamefont {J.~K.}\ \bibnamefont
  {Jang}}, \bibinfo {author} {\bibfnamefont {M.}~\bibnamefont {Erkintalo}},
  \bibinfo {author} {\bibfnamefont {S.~G.}\ \bibnamefont {Murdoch}}, \ and\
  \bibinfo {author} {\bibfnamefont {S.}~\bibnamefont {Coen}},\ }\bibfield
  {title} {\enquote {\bibinfo {title} {Observation of dispersive wave emission
  by temporal cavity solitons},}\ }\href@noop {} {\bibfield  {journal}
  {\bibinfo  {journal} {Optics Letters}\ }\textbf {\bibinfo {volume} {39}},\
  \bibinfo {pages} {5503} (\bibinfo {year} {2014})}\BibitemShut {NoStop}%
\bibitem [{\citenamefont {Englebert}\ \emph {et~al.}(2021)\citenamefont
  {Englebert}, \citenamefont {Mas~Arabí}, \citenamefont {Parra-Rivas},
  \citenamefont {Gorza},\ and\ \citenamefont {Leo}}]{englebert_temporal_2021}%
  \BibitemOpen
  \bibfield  {author} {\bibinfo {author} {\bibfnamefont {N.}~\bibnamefont
  {Englebert}}, \bibinfo {author} {\bibfnamefont {C.}~\bibnamefont
  {Mas~Arabí}}, \bibinfo {author} {\bibfnamefont {P.}~\bibnamefont
  {Parra-Rivas}}, \bibinfo {author} {\bibfnamefont {S.-P.}\ \bibnamefont
  {Gorza}}, \ and\ \bibinfo {author} {\bibfnamefont {F.}~\bibnamefont {Leo}},\
  }\bibfield  {title} {\enquote {\bibinfo {title} {Temporal solitons in a
  coherently driven active resonator},}\ }\href@noop {} {\bibfield  {journal}
  {\bibinfo  {journal} {Nature Photonics}\ }\textbf {\bibinfo {volume} {15}},\
  \bibinfo {pages} {536--541} (\bibinfo {year} {2021})}\BibitemShut {NoStop}%
\bibitem [{\citenamefont {Jia}\ \emph {et~al.}(2020)\citenamefont {Jia},
  \citenamefont {Wang}, \citenamefont {Kwon}, \citenamefont {Wang},
  \citenamefont {Tsao}, \citenamefont {Liu}, \citenamefont {Ni}, \citenamefont
  {Guo}, \citenamefont {Yang}, \citenamefont {Jiang}, \citenamefont {Kim},
  \citenamefont {Zhu}, \citenamefont {Xie},\ and\ \citenamefont
  {Huang}}]{jia_photonic_2020}%
  \BibitemOpen
  \bibfield  {author} {\bibinfo {author} {\bibfnamefont {K.}~\bibnamefont
  {Jia}}, \bibinfo {author} {\bibfnamefont {X.}~\bibnamefont {Wang}}, \bibinfo
  {author} {\bibfnamefont {D.}~\bibnamefont {Kwon}}, \bibinfo {author}
  {\bibfnamefont {J.}~\bibnamefont {Wang}}, \bibinfo {author} {\bibfnamefont
  {E.}~\bibnamefont {Tsao}}, \bibinfo {author} {\bibfnamefont {H.}~\bibnamefont
  {Liu}}, \bibinfo {author} {\bibfnamefont {X.}~\bibnamefont {Ni}}, \bibinfo
  {author} {\bibfnamefont {J.}~\bibnamefont {Guo}}, \bibinfo {author}
  {\bibfnamefont {M.}~\bibnamefont {Yang}}, \bibinfo {author} {\bibfnamefont
  {X.}~\bibnamefont {Jiang}}, \bibinfo {author} {\bibfnamefont
  {J.}~\bibnamefont {Kim}}, \bibinfo {author} {\bibfnamefont {S.-n.}\
  \bibnamefont {Zhu}}, \bibinfo {author} {\bibfnamefont {Z.}~\bibnamefont
  {Xie}}, \ and\ \bibinfo {author} {\bibfnamefont {S.-W.}\ \bibnamefont
  {Huang}},\ }\bibfield  {title} {\enquote {\bibinfo {title} {Photonic
  {Flywheel} in a {Monolithic} {Fiber} {Resonator}},}\ }\href@noop {}
  {\bibfield  {journal} {\bibinfo  {journal} {Physical Review Letters}\
  }\textbf {\bibinfo {volume} {125}},\ \bibinfo {pages} {143902} (\bibinfo
  {year} {2020})}\BibitemShut {NoStop}%
\bibitem [{\citenamefont {Xiao}\ \emph {et~al.}(2023)\citenamefont {Xiao},
  \citenamefont {Li}, \citenamefont {Cai}, \citenamefont {Zhang}, \citenamefont
  {Huang}, \citenamefont {Li}, \citenamefont {Yao}, \citenamefont {Wu},\ and\
  \citenamefont {Chen}}]{xiao_near-zero-dispersion_2023}%
  \BibitemOpen
  \bibfield  {author} {\bibinfo {author} {\bibfnamefont {Z.}~\bibnamefont
  {Xiao}}, \bibinfo {author} {\bibfnamefont {T.}~\bibnamefont {Li}}, \bibinfo
  {author} {\bibfnamefont {M.}~\bibnamefont {Cai}}, \bibinfo {author}
  {\bibfnamefont {H.}~\bibnamefont {Zhang}}, \bibinfo {author} {\bibfnamefont
  {Y.}~\bibnamefont {Huang}}, \bibinfo {author} {\bibfnamefont
  {C.}~\bibnamefont {Li}}, \bibinfo {author} {\bibfnamefont {B.}~\bibnamefont
  {Yao}}, \bibinfo {author} {\bibfnamefont {K.}~\bibnamefont {Wu}}, \ and\
  \bibinfo {author} {\bibfnamefont {J.}~\bibnamefont {Chen}},\ }\bibfield
  {title} {\enquote {\bibinfo {title} {Near-zero-dispersion soliton and
  broadband modulational instability {Kerr} microcombs in anomalous
  dispersion},}\ }\href@noop {} {\bibfield  {journal} {\bibinfo  {journal}
  {Light: Science \& Applications}\ }\textbf {\bibinfo {volume} {12}},\
  \bibinfo {pages} {33} (\bibinfo {year} {2023})}\BibitemShut {NoStop}%
\bibitem [{\citenamefont {Bunel}\ \emph
  {et~al.}(2023{\natexlab{a}})\citenamefont {Bunel}, \citenamefont {Conforti},
  \citenamefont {Ziani}, \citenamefont {Lumeau}, \citenamefont {Moreau},
  \citenamefont {Fernandez}, \citenamefont {Llopis}, \citenamefont {Roul},
  \citenamefont {Perego}, \citenamefont {Wong} \emph
  {et~al.}}]{bunel2023observation}%
  \BibitemOpen
  \bibfield  {author} {\bibinfo {author} {\bibfnamefont {T.}~\bibnamefont
  {Bunel}}, \bibinfo {author} {\bibfnamefont {M.}~\bibnamefont {Conforti}},
  \bibinfo {author} {\bibfnamefont {Z.}~\bibnamefont {Ziani}}, \bibinfo
  {author} {\bibfnamefont {J.}~\bibnamefont {Lumeau}}, \bibinfo {author}
  {\bibfnamefont {A.}~\bibnamefont {Moreau}}, \bibinfo {author} {\bibfnamefont
  {A.}~\bibnamefont {Fernandez}}, \bibinfo {author} {\bibfnamefont
  {O.}~\bibnamefont {Llopis}}, \bibinfo {author} {\bibfnamefont
  {J.}~\bibnamefont {Roul}}, \bibinfo {author} {\bibfnamefont {A.~M.}\
  \bibnamefont {Perego}}, \bibinfo {author} {\bibfnamefont {K.~K.}\
  \bibnamefont {Wong}},  \emph {et~al.},\ }\bibfield  {title} {\enquote
  {\bibinfo {title} {Observation of modulation instability kerr frequency combs
  in a fiber fabry--p{\'e}rot resonator},}\ }\href@noop {} {\bibfield
  {journal} {\bibinfo  {journal} {Optics Letters}\ }\textbf {\bibinfo {volume}
  {48}},\ \bibinfo {pages} {275--278} (\bibinfo {year}
  {2023}{\natexlab{a}})}\BibitemShut {NoStop}%
\bibitem [{\citenamefont {Li}\ \emph {et~al.}(2022)\citenamefont {Li},
  \citenamefont {Xu}, \citenamefont {Shamailov}, \citenamefont {Wen},
  \citenamefont {Wang}, \citenamefont {Wei}, \citenamefont {Yang},
  \citenamefont {Coen}, \citenamefont {Murdoch},\ and\ \citenamefont
  {Erkintalo}}]{li_ultrashort_2022}%
  \BibitemOpen
  \bibfield  {author} {\bibinfo {author} {\bibfnamefont {Z.}~\bibnamefont
  {Li}}, \bibinfo {author} {\bibfnamefont {Y.}~\bibnamefont {Xu}}, \bibinfo
  {author} {\bibfnamefont {S.}~\bibnamefont {Shamailov}}, \bibinfo {author}
  {\bibfnamefont {X.}~\bibnamefont {Wen}}, \bibinfo {author} {\bibfnamefont
  {W.}~\bibnamefont {Wang}}, \bibinfo {author} {\bibfnamefont {X.}~\bibnamefont
  {Wei}}, \bibinfo {author} {\bibfnamefont {Z.}~\bibnamefont {Yang}}, \bibinfo
  {author} {\bibfnamefont {S.}~\bibnamefont {Coen}}, \bibinfo {author}
  {\bibfnamefont {S.~G.}\ \bibnamefont {Murdoch}}, \ and\ \bibinfo {author}
  {\bibfnamefont {M.}~\bibnamefont {Erkintalo}},\ }\href
  {http://arxiv.org/abs/2212.08223} {\enquote {\bibinfo {title} {Ultrashort
  dissipative {Raman} solitons in {Kerr} resonators driven with phase-coherent
  optical pulses},}\ } (\bibinfo {year} {2022}),\ \bibinfo {note}
  {arXiv:2212.08223 [nlin, physics:physics]},\ \Eprint
  {http://arxiv.org/abs/2212.08223} {2212.08223} \BibitemShut {NoStop}%
\bibitem [{\citenamefont {Bunel}\ \emph
  {et~al.}(2023{\natexlab{b}})\citenamefont {Bunel}, \citenamefont {Conforti},
  \citenamefont {Lumeau}, \citenamefont {Moreau}, \citenamefont {Fernandez},
  \citenamefont {Llopis}, \citenamefont {Roul}, \citenamefont {Perego},\ and\
  \citenamefont {Mussot}}]{postdeadline}%
  \BibitemOpen
  \bibfield  {author} {\bibinfo {author} {\bibfnamefont {T.}~\bibnamefont
  {Bunel}}, \bibinfo {author} {\bibfnamefont {M.}~\bibnamefont {Conforti}},
  \bibinfo {author} {\bibfnamefont {J.}~\bibnamefont {Lumeau}}, \bibinfo
  {author} {\bibfnamefont {A.}~\bibnamefont {Moreau}}, \bibinfo {author}
  {\bibfnamefont {A.}~\bibnamefont {Fernandez}}, \bibinfo {author}
  {\bibfnamefont {O.}~\bibnamefont {Llopis}}, \bibinfo {author} {\bibfnamefont
  {J.}~\bibnamefont {Roul}}, \bibinfo {author} {\bibfnamefont {A.~M.}\
  \bibnamefont {Perego}}, \ and\ \bibinfo {author} {\bibfnamefont
  {A.}~\bibnamefont {Mussot}},\ }\bibfield  {title} {\enquote {\bibinfo {title}
  {Unexpected phase-locked brillouin kerr frequency comb in fiber fabry perot
  resonators},}\ \ }(\bibinfo  {publisher} {CLEO{\textregistered}/Europe 2023 -
  Postdeadline session},\ \bibinfo {year} {2023})\BibitemShut {NoStop}%
\bibitem [{\citenamefont {Nie}\ \emph {et~al.}(2022)\citenamefont {Nie},
  \citenamefont {Jia}, \citenamefont {Xie}, \citenamefont {Zhu}, \citenamefont
  {Xie},\ and\ \citenamefont {Huang}}]{nie_synthesized_2022}%
  \BibitemOpen
  \bibfield  {author} {\bibinfo {author} {\bibfnamefont {M.}~\bibnamefont
  {Nie}}, \bibinfo {author} {\bibfnamefont {K.}~\bibnamefont {Jia}}, \bibinfo
  {author} {\bibfnamefont {Y.}~\bibnamefont {Xie}}, \bibinfo {author}
  {\bibfnamefont {S.}~\bibnamefont {Zhu}}, \bibinfo {author} {\bibfnamefont
  {Z.}~\bibnamefont {Xie}}, \ and\ \bibinfo {author} {\bibfnamefont {S.-W.}\
  \bibnamefont {Huang}},\ }\bibfield  {title} {\enquote {\bibinfo {title}
  {Synthesized spatiotemporal mode-locking and photonic flywheel in multimode
  mesoresonators},}\ }\href@noop {} {\bibfield  {journal} {\bibinfo  {journal}
  {Nature Communications}\ }\textbf {\bibinfo {volume} {13}},\ \bibinfo {pages}
  {6395} (\bibinfo {year} {2022})}\BibitemShut {NoStop}%
\bibitem [{\citenamefont {Zideluns}\ \emph {et~al.}(2021)\citenamefont
  {Zideluns}, \citenamefont {Lemarchand}, \citenamefont {Arhilger},
  \citenamefont {Hagedorn},\ and\ \citenamefont
  {Lumeau}}]{zideluns_automated_2021}%
  \BibitemOpen
  \bibfield  {author} {\bibinfo {author} {\bibfnamefont {J.}~\bibnamefont
  {Zideluns}}, \bibinfo {author} {\bibfnamefont {F.}~\bibnamefont
  {Lemarchand}}, \bibinfo {author} {\bibfnamefont {D.}~\bibnamefont
  {Arhilger}}, \bibinfo {author} {\bibfnamefont {H.}~\bibnamefont {Hagedorn}},
  \ and\ \bibinfo {author} {\bibfnamefont {J.}~\bibnamefont {Lumeau}},\
  }\bibfield  {title} {\enquote {\bibinfo {title} {Automated optical monitoring
  wavelength selection for thin-film filters},}\ }\href@noop {} {\bibfield
  {journal} {\bibinfo  {journal} {Optics Express}\ }\textbf {\bibinfo {volume}
  {29}},\ \bibinfo {pages} {33398} (\bibinfo {year} {2021})}\BibitemShut
  {NoStop}%
\bibitem [{\citenamefont {Nishimoto}\ \emph {et~al.}(2022)\citenamefont
  {Nishimoto}, \citenamefont {Minoshima}, \citenamefont {Yasui},\ and\
  \citenamefont {Kuse}}]{nishimoto_thermal_2022}%
  \BibitemOpen
  \bibfield  {author} {\bibinfo {author} {\bibfnamefont {K.}~\bibnamefont
  {Nishimoto}}, \bibinfo {author} {\bibfnamefont {K.}~\bibnamefont
  {Minoshima}}, \bibinfo {author} {\bibfnamefont {T.}~\bibnamefont {Yasui}}, \
  and\ \bibinfo {author} {\bibfnamefont {N.}~\bibnamefont {Kuse}},\ }\bibfield
  {title} {\enquote {\bibinfo {title} {Thermal control of a {Kerr}
  microresonator soliton comb via an optical sideband},}\ }\href@noop {}
  {\bibfield  {journal} {\bibinfo  {journal} {Optics Letters}\ }\textbf
  {\bibinfo {volume} {47}},\ \bibinfo {pages} {281} (\bibinfo {year}
  {2022})}\BibitemShut {NoStop}%
\bibitem [{\citenamefont {Drever}\ \emph {et~al.}(1983)\citenamefont {Drever},
  \citenamefont {Hall}, \citenamefont {Kowalski}, \citenamefont {Hough},
  \citenamefont {Ford}, \citenamefont {Munley},\ and\ \citenamefont
  {Ward}}]{drever_laser_1983}%
  \BibitemOpen
  \bibfield  {author} {\bibinfo {author} {\bibfnamefont {R.~W.~P.}\
  \bibnamefont {Drever}}, \bibinfo {author} {\bibfnamefont {J.~L.}\
  \bibnamefont {Hall}}, \bibinfo {author} {\bibfnamefont {F.~V.}\ \bibnamefont
  {Kowalski}}, \bibinfo {author} {\bibfnamefont {J.}~\bibnamefont {Hough}},
  \bibinfo {author} {\bibfnamefont {G.~M.}\ \bibnamefont {Ford}}, \bibinfo
  {author} {\bibfnamefont {A.~J.}\ \bibnamefont {Munley}}, \ and\ \bibinfo
  {author} {\bibfnamefont {H.}~\bibnamefont {Ward}},\ }\bibfield  {title}
  {\enquote {\bibinfo {title} {Laser phase and frequency stabilization using an
  optical resonator},}\ }\href@noop {} {\bibfield  {journal} {\bibinfo
  {journal} {Applied Physics B Photophysics and Laser Chemistry}\ }\textbf
  {\bibinfo {volume} {31}},\ \bibinfo {pages} {97--105} (\bibinfo {year}
  {1983})}\BibitemShut {NoStop}%
\bibitem [{\citenamefont {Black}(2001)}]{black_introduction_2001}%
  \BibitemOpen
  \bibfield  {author} {\bibinfo {author} {\bibfnamefont {E.~D.}\ \bibnamefont
  {Black}},\ }\bibfield  {title} {\enquote {\bibinfo {title} {An introduction
  to {Pound}–{Drever}–{Hall} laser frequency stabilization},}\ }\href@noop
  {} {\bibfield  {journal} {\bibinfo  {journal} {American Journal of Physics}\
  }\textbf {\bibinfo {volume} {69}},\ \bibinfo {pages} {79--87} (\bibinfo
  {year} {2001})}\BibitemShut {NoStop}%
\bibitem [{\citenamefont {Coen}\ and\ \citenamefont
  {Erkintalo}(2013)}]{coen_universal_2013}%
  \BibitemOpen
  \bibfield  {author} {\bibinfo {author} {\bibfnamefont {S.}~\bibnamefont
  {Coen}}\ and\ \bibinfo {author} {\bibfnamefont {M.}~\bibnamefont
  {Erkintalo}},\ }\bibfield  {title} {\enquote {\bibinfo {title} {Universal
  scaling laws of {Kerr} frequency combs},}\ }\href@noop {} {\bibfield
  {journal} {\bibinfo  {journal} {Optics Letters}\ }\textbf {\bibinfo {volume}
  {38}},\ \bibinfo {pages} {1790} (\bibinfo {year} {2013})}\BibitemShut
  {NoStop}%
\bibitem [{\citenamefont {Parra-Rivas}\ \emph {et~al.}(2014)\citenamefont
  {Parra-Rivas}, \citenamefont {Gomila}, \citenamefont {Matías}, \citenamefont
  {Coen},\ and\ \citenamefont {Gelens}}]{parra-rivas_dynamics_2014}%
  \BibitemOpen
  \bibfield  {author} {\bibinfo {author} {\bibfnamefont {P.}~\bibnamefont
  {Parra-Rivas}}, \bibinfo {author} {\bibfnamefont {D.}~\bibnamefont {Gomila}},
  \bibinfo {author} {\bibfnamefont {M.~A.}\ \bibnamefont {Matías}}, \bibinfo
  {author} {\bibfnamefont {S.}~\bibnamefont {Coen}}, \ and\ \bibinfo {author}
  {\bibfnamefont {L.}~\bibnamefont {Gelens}},\ }\bibfield  {title} {\enquote
  {\bibinfo {title} {Dynamics of localized and patterned structures in the
  {Lugiato}-{Lefever} equation determine the stability and shape of optical
  frequency combs},}\ }\href@noop {} {\bibfield  {journal} {\bibinfo  {journal}
  {Physical Review A}\ }\textbf {\bibinfo {volume} {89}},\ \bibinfo {pages}
  {043813} (\bibinfo {year} {2014})}\BibitemShut {NoStop}%
\bibitem [{\citenamefont {Cole}\ \emph {et~al.}(2018)\citenamefont {Cole},
  \citenamefont {Gatti}, \citenamefont {Papp}, \citenamefont {Prati},\ and\
  \citenamefont {Lugiato}}]{cole_theory_2018}%
  \BibitemOpen
  \bibfield  {author} {\bibinfo {author} {\bibfnamefont {D.~C.}\ \bibnamefont
  {Cole}}, \bibinfo {author} {\bibfnamefont {A.}~\bibnamefont {Gatti}},
  \bibinfo {author} {\bibfnamefont {S.~B.}\ \bibnamefont {Papp}}, \bibinfo
  {author} {\bibfnamefont {F.}~\bibnamefont {Prati}}, \ and\ \bibinfo {author}
  {\bibfnamefont {L.}~\bibnamefont {Lugiato}},\ }\bibfield  {title} {\enquote
  {\bibinfo {title} {Theory of kerr frequency combs in fabry-perot
  resonators},}\ }\href@noop {} {\bibfield  {journal} {\bibinfo  {journal}
  {Physical Review A}\ }\textbf {\bibinfo {volume} {98}},\ \bibinfo {pages}
  {013831} (\bibinfo {year} {2018})}\BibitemShut {NoStop}%
\bibitem [{\citenamefont {Campbell}\ \emph {et~al.}(2023)\citenamefont
  {Campbell}, \citenamefont {Hill}, \citenamefont {Del'Haye},\ and\
  \citenamefont {Oppo}}]{Campbell2023}%
  \BibitemOpen
  \bibfield  {author} {\bibinfo {author} {\bibfnamefont {G.~N.}\ \bibnamefont
  {Campbell}}, \bibinfo {author} {\bibfnamefont {L.}~\bibnamefont {Hill}},
  \bibinfo {author} {\bibfnamefont {P.}~\bibnamefont {Del'Haye}}, \ and\
  \bibinfo {author} {\bibfnamefont {G.-L.}\ \bibnamefont {Oppo}},\ }\bibfield
  {title} {\enquote {\bibinfo {title} {Dark solitons in fabry-p\'erot
  resonators with kerr media and normal dispersion},}\ }\href {\doibase
  10.1103/PhysRevA.108.033505} {\bibfield  {journal} {\bibinfo  {journal}
  {Phys. Rev. A}\ }\textbf {\bibinfo {volume} {108}},\ \bibinfo {pages}
  {033505} (\bibinfo {year} {2023})}\BibitemShut {NoStop}%
\bibitem [{\citenamefont {Firth}(1981)}]{firth_stability_1981}%
  \BibitemOpen
  \bibfield  {author} {\bibinfo {author} {\bibfnamefont {W.}~\bibnamefont
  {Firth}},\ }\bibfield  {title} {\enquote {\bibinfo {title} {Stability of
  nonlinear {Fabry}-{Perot} resonators},}\ }\href {\doibase
  10.1016/0030-4018(81)90106-1} {\bibfield  {journal} {\bibinfo  {journal}
  {Optics Communications}\ }\textbf {\bibinfo {volume} {39}},\ \bibinfo {pages}
  {343--346} (\bibinfo {year} {1981})}\BibitemShut {NoStop}%
\bibitem [{\citenamefont {Firth}\ \emph {et~al.}(2021)\citenamefont {Firth},
  \citenamefont {Geddes}, \citenamefont {Karst},\ and\ \citenamefont
  {Oppo}}]{firth_analytic_2021}%
  \BibitemOpen
  \bibfield  {author} {\bibinfo {author} {\bibfnamefont {W.~J.}\ \bibnamefont
  {Firth}}, \bibinfo {author} {\bibfnamefont {J.~B.}\ \bibnamefont {Geddes}},
  \bibinfo {author} {\bibfnamefont {N.~J.}\ \bibnamefont {Karst}}, \ and\
  \bibinfo {author} {\bibfnamefont {G.-L.}\ \bibnamefont {Oppo}},\ }\bibfield
  {title} {\enquote {\bibinfo {title} {Analytic instability thresholds in
  folded kerr resonators of arbitrary finesse},}\ }\href@noop {} {\bibfield
  {journal} {\bibinfo  {journal} {Physical Review A}\ }\textbf {\bibinfo
  {volume} {103}},\ \bibinfo {pages} {023510} (\bibinfo {year}
  {2021})}\BibitemShut {NoStop}%
\bibitem [{\citenamefont {Ziani}\ \emph {et~al.}(2023)\citenamefont {Ziani},
  \citenamefont {Bunel}, \citenamefont {Perego}, \citenamefont {Mussot},\ and\
  \citenamefont {Conforti}}]{ziani_theory_2023}%
  \BibitemOpen
  \bibfield  {author} {\bibinfo {author} {\bibfnamefont {Z.}~\bibnamefont
  {Ziani}}, \bibinfo {author} {\bibfnamefont {T.}~\bibnamefont {Bunel}},
  \bibinfo {author} {\bibfnamefont {A.~M.}\ \bibnamefont {Perego}}, \bibinfo
  {author} {\bibfnamefont {A.}~\bibnamefont {Mussot}}, \ and\ \bibinfo {author}
  {\bibfnamefont {M.}~\bibnamefont {Conforti}},\ }\href@noop {} {\enquote
  {\bibinfo {title} {Theory of modulation instability in {Kerr} {Fabry}-{Perot}
  resonators beyond the mean field limit},}\ } (\bibinfo {year} {2023}),\
  \bibinfo {note} {arXiv:2307.13488 [nlin, physics:physics]}\BibitemShut
  {NoStop}%
\bibitem [{\citenamefont {Yi}\ \emph {et~al.}(2015)\citenamefont {Yi},
  \citenamefont {Yang}, \citenamefont {Yang}, \citenamefont {Suh},\ and\
  \citenamefont {Vahala}}]{yi_soliton_2015}%
  \BibitemOpen
  \bibfield  {author} {\bibinfo {author} {\bibfnamefont {X.}~\bibnamefont
  {Yi}}, \bibinfo {author} {\bibfnamefont {Q.-F.}\ \bibnamefont {Yang}},
  \bibinfo {author} {\bibfnamefont {K.~Y.}\ \bibnamefont {Yang}}, \bibinfo
  {author} {\bibfnamefont {M.-G.}\ \bibnamefont {Suh}}, \ and\ \bibinfo
  {author} {\bibfnamefont {K.}~\bibnamefont {Vahala}},\ }\bibfield  {title}
  {\enquote {\bibinfo {title} {Soliton frequency comb at microwave rates in a
  high-{Q} silica microresonator},}\ }\href@noop {} {\bibfield  {journal}
  {\bibinfo  {journal} {Optica}\ }\textbf {\bibinfo {volume} {2}},\ \bibinfo
  {pages} {1078} (\bibinfo {year} {2015})}\BibitemShut {NoStop}%
\bibitem [{\citenamefont {Conforti}\ and\ \citenamefont
  {Trillo}(2013)}]{conforti_dispersive_2013}%
  \BibitemOpen
  \bibfield  {author} {\bibinfo {author} {\bibfnamefont {M.}~\bibnamefont
  {Conforti}}\ and\ \bibinfo {author} {\bibfnamefont {S.}~\bibnamefont
  {Trillo}},\ }\bibfield  {title} {\enquote {\bibinfo {title} {Dispersive wave
  emission from wave breaking},}\ }\href@noop {} {\bibfield  {journal}
  {\bibinfo  {journal} {Optics Letters}\ }\textbf {\bibinfo {volume} {38}},\
  \bibinfo {pages} {3815} (\bibinfo {year} {2013})}\BibitemShut {NoStop}%
\bibitem [{\citenamefont {Erkintalo}\ \emph {et~al.}(2012)\citenamefont
  {Erkintalo}, \citenamefont {Xu}, \citenamefont {Murdoch}, \citenamefont
  {Dudley},\ and\ \citenamefont {Genty}}]{erkintalo_cascaded_2012}%
  \BibitemOpen
  \bibfield  {author} {\bibinfo {author} {\bibfnamefont {M.}~\bibnamefont
  {Erkintalo}}, \bibinfo {author} {\bibfnamefont {Y.~Q.}\ \bibnamefont {Xu}},
  \bibinfo {author} {\bibfnamefont {S.~G.}\ \bibnamefont {Murdoch}}, \bibinfo
  {author} {\bibfnamefont {J.~M.}\ \bibnamefont {Dudley}}, \ and\ \bibinfo
  {author} {\bibfnamefont {G.}~\bibnamefont {Genty}},\ }\bibfield  {title}
  {\enquote {\bibinfo {title} {Cascaded {Phase} {Matching} and {Nonlinear}
  {Symmetry} {Breaking} in {Fiber} {Frequency} {Combs}},}\ }\href@noop {}
  {\bibfield  {journal} {\bibinfo  {journal} {Physical Review Letters}\
  }\textbf {\bibinfo {volume} {109}},\ \bibinfo {pages} {223904} (\bibinfo
  {year} {2012})}\BibitemShut {NoStop}%
\bibitem [{\citenamefont {Milián}\ and\ \citenamefont
  {Skryabin}(2014)}]{milian_soliton_2014}%
  \BibitemOpen
  \bibfield  {author} {\bibinfo {author} {\bibfnamefont {C.}~\bibnamefont
  {Milián}}\ and\ \bibinfo {author} {\bibfnamefont {D.}~\bibnamefont
  {Skryabin}},\ }\bibfield  {title} {\enquote {\bibinfo {title} {Soliton
  families and resonant radiation in a micro-ring resonator near zero
  group-velocity dispersion},}\ }\href@noop {} {\bibfield  {journal} {\bibinfo
  {journal} {Optics Express}\ }\textbf {\bibinfo {volume} {22}},\ \bibinfo
  {pages} {3732} (\bibinfo {year} {2014})}\BibitemShut {NoStop}%
\bibitem [{\citenamefont {Brasch}\ \emph {et~al.}(2016)\citenamefont {Brasch},
  \citenamefont {Geiselmann}, \citenamefont {Herr}, \citenamefont {Lihachev},
  \citenamefont {Pfeiffer}, \citenamefont {Gorodetsky},\ and\ \citenamefont
  {Kippenberg}}]{brasch_photonic_2016}%
  \BibitemOpen
  \bibfield  {author} {\bibinfo {author} {\bibfnamefont {V.}~\bibnamefont
  {Brasch}}, \bibinfo {author} {\bibfnamefont {M.}~\bibnamefont {Geiselmann}},
  \bibinfo {author} {\bibfnamefont {T.}~\bibnamefont {Herr}}, \bibinfo {author}
  {\bibfnamefont {G.}~\bibnamefont {Lihachev}}, \bibinfo {author}
  {\bibfnamefont {M.~H.~P.}\ \bibnamefont {Pfeiffer}}, \bibinfo {author}
  {\bibfnamefont {M.~L.}\ \bibnamefont {Gorodetsky}}, \ and\ \bibinfo {author}
  {\bibfnamefont {T.~J.}\ \bibnamefont {Kippenberg}},\ }\bibfield  {title}
  {\enquote {\bibinfo {title} {Photonic chip–based optical frequency comb
  using soliton {Cherenkov} radiation},}\ }\href@noop {} {\bibfield  {journal}
  {\bibinfo  {journal} {Science}\ }\textbf {\bibinfo {volume} {351}},\ \bibinfo
  {pages} {357--360} (\bibinfo {year} {2016})}\BibitemShut {NoStop}%
\bibitem [{\citenamefont {Wildi}\ \emph {et~al.}(2023)\citenamefont {Wildi},
  \citenamefont {Gaafar}, \citenamefont {Voumard}, \citenamefont {Ludwig},\
  and\ \citenamefont {Herr}}]{wildi_dissipative_2023}%
  \BibitemOpen
  \bibfield  {author} {\bibinfo {author} {\bibfnamefont {T.}~\bibnamefont
  {Wildi}}, \bibinfo {author} {\bibfnamefont {M.~A.}\ \bibnamefont {Gaafar}},
  \bibinfo {author} {\bibfnamefont {T.}~\bibnamefont {Voumard}}, \bibinfo
  {author} {\bibfnamefont {M.}~\bibnamefont {Ludwig}}, \ and\ \bibinfo {author}
  {\bibfnamefont {T.}~\bibnamefont {Herr}},\ }\bibfield  {title} {\enquote
  {\bibinfo {title} {Dissipative {Kerr} solitons in integrated
  {Fabry}–{Perot} microresonators},}\ }\href@noop {} {\bibfield  {journal}
  {\bibinfo  {journal} {Optica}\ }\textbf {\bibinfo {volume} {10}},\ \bibinfo
  {pages} {650} (\bibinfo {year} {2023})}\BibitemShut {NoStop}%
\end{thebibliography}%

\end{document}